\documentclass[twocolumn,secnumarabic,amssymb, amsmath, nofootinbib,tightenlines,
nobibnotes, aps, prl,epsfig]{revtex4}
\usepackage{graphicx}
\usepackage{dcolumn}
\usepackage{bm}
\begin{document}
\preprint{APS/123-QED}
\title{The behavior of the structure function by using the effective exponent at low-$x$}
\author{B.Rezaei}%
\email{brezaei@razi.ac.ir }
\author{ G.R.Boroun}%
 \email{grboroun@gmail.com; boroun@razi.ac.ir }
\affiliation{ Physics Department, Razi University, Kermanshah
67149, Iran}
\date{\today}
\begin{abstract}
An analytical solution of the QCD evolution equations for the
singlet and gluon distribution is presented. We decouple DGLAP
evolution equations into the initial conditions by using a Laplace
transform method  at $N^{n}LO$ analysis. The relationship between
the nonlinear behavior and color dipole model is considered based
on an effective exponent behavior at low-$x$ values. We obtain the
effective exponent at NLO analysis from the decoupled behavior of
the distribution functions. The proton structure function compared
with H1 data from the inclusive structure function
$F_{2}(x,Q^{2})$ for $x
{\leq}~ 10^{-2}$ and $5 {\leq} Q^{2} {\leq} 250~GeV^{2}$.\\
\end{abstract}

\maketitle
\tableofcontents

\section{1. Introduction}

 Parton distribution functions can be
used as fundamental tools to extract the structure functions of
proton in deep inelastic scattering (DIS) processes. Deep
inelastic scattering is characterised by the variables $Q^{2}$ and
$x$ where $Q^{2}$ is  the virtuality of the exchanged virtual
photon and $x$ is the fraction of proton momentum carried by the
parton. These distributions prescribed in Quantum Chromodynamics
(QCD) and extracted by the DGLAP [1] evolution equations. The
solutions of these evolution equations allows us to predict the
gluon and sea quark distributions at low values of $x$ for
understanding the quark-gluon dynamics inside the nucleon. On the
basis of the DGLAP evolution equations  at low $x$ it is known
that the dominate source for distribution functions is the gluon
density. It is expected that the gluon density can not grow
forever due to Froissart bound at very low values of $x$. The
gluon density behavior tamed in this region due to the correlative
interactions between the gluons.\\
 This behavior of the gluon
density will be checked at the Large Hadron electron Collider
(LHeC) where leads to beyond a $\mathrm{TeV}$ in center-of-mass
energy [2]. The LHeC center of mass energy is $1.3~\mathrm{TeV}$
where extends the kinematic ranges in $x$ and $Q^{2}$ by factors
of $\sim~20$ than those accessible at HERA. The  DIS kinematics
reaches $\simeq 1~TeV^{2}$ and $\simeq 10^{-6}$ for $Q^{2}$ and
$x$ respectively where the gluon distribution has a non-linear
behavior in this region. Indeed the LHeC is designed for study
Higgs boson, Top
quark production and CFT-ADS correspondence [3].\\
 The structure
functions  are sensitive to the gluon saturation in the LHeC
kinematic range. Therefore the dynamics of gluon behavior is an
interesting subject in this region. The nonlinear behavior is
important when $3\pi \alpha_{s}G(x,Q^{2}){\geq}Q^{2}R^{2}$, where
$R$ is the size of the target. In such a case we reach the region
of high gluon density QCD which annihilation of gluons (introduced
by the vertex $\textit{gluon+gluon}$$\rightarrow$$\textit{gluon}$)
becomes important [4]. In fact the gluon recombination terms lead
to the nonlinear corrections. These multiple gluon-gluon
interactions provide nonlinear corrections to the linear DGLAP
evolution equations. These nonlinear corrections have been
calculated by Gribov, Levin,Ryskin, Mueller and Qiu (GLR-MQ) [5]
and
tame the parton distributions behavior at sufficiently low $x$.\\
In recent years [6-7], the Regge-like behavior for the gluon
density used in the GLR-MQ equation. The key ingredient is the
gluon behavior in this region, especially the effect of screening
on the Regee trajectory. In Ref.[8] the general behavior of the
gluon density is studied at leading order using GLR-MQ equation
with respect to the Laplace transformation method. This behavior
tamed by screening effects. This leads to the reduction of the
growth of parton distributions, which is called parton saturation.
Saturation is known by the  scale $Q_{s}^{2}(x)$ where the
nonlinear effects appear for $Q^{2}<Q_{s}^{2}$. Here function
$Q_{s}$- called saturation scale- was taken in the following form
$Q^{2}_{s}=Q^{2}_{0}(\frac{x}{x_{0}})^{-\lambda}$ where $Q_{0}$
and $x_{0}$ are free parameters which can be extracted from the
data and exponent $\lambda$ is a dynamical quantity. We
concentrate on the nonlinear behavior
 by the saturation model based on the decoupled solutions for the gluon and
 singlet
distribution functions.\\
The paper is organized as follows. In Section 2, the QCD coupled
DGLAP evolution equations studied and presented an analytical
solution for the decoupled DGLAP evolution equation for the parton
distribution functions (PDFs)  based on the Laplace transform
method. In Section 3 we apply the nonlinear behavior to the
decoupled DGLAP equations and introduce a transition to the
saturation model at low-$x$ values. Section 4 is devoted to the
results for the gluon distribution function and  proton structure
function. The effective exponents into the behavior of the
distribution functions are presented. The effective exponents for
HERA data are obtained in Section 5 in accordance with the
decoupled solutions. The behavior of the structure function is
compared with H1 data for $x {\leq}~ 10^{-2}$ and $5 {\leq} Q^{2}
{\leq} 250~GeV^{2}$ in Section 6. Finally we give our summary and
conclusions in section
7.\\
\section{2. Formulism Decoupling DGLAP}
The study of linear DGLAP evolution equations by using a Laplace
transform  have a history and many applications [9], both for
 coupled as well as decoupled distribution functions. In Ref.[9], a general method has been derived for calculation the
evolution of parton distribution functions within the Laplace
transform method. The polarized and unpolarized DGLAP equations
for the QCD evolution of parton distribution functions
demonstrated in
 Refs.[10-12].\\
Starting from the coupled DGLAP evolution equations by
  using the Laplace transform method. These equations can be directly calculated as a convolution of
the $s$-space with impact factors that encode the splitting
functions in that process, as we have
 \begin{eqnarray}
\frac{\partial f_{s}(s,t)}{\partial
t}&=&(\sum_{n=1}a^{n}(t)\Phi_{f}^{(n)}(s))f_{s}(s,t)\nonumber\\
&&+(\sum_{n=1}a^{n}(t)\Theta_{f}^{(n)}(s))g(s,t),\\
\mathrm{and}\nonumber\\
 \frac{\partial
g(s,t)}{\partial
t}&=&(\sum_{n=1}a^{n}(t)\Phi_{g}^{(n)}(s))f_{s}(s,t)\nonumber\\
&&+(\sum_{n=1}a^{n}(t)\Theta_{g}^{(n)}(s))g(s,t).
 \end{eqnarray}
 The running coupling constant in the high-loop corrections
of above evolution equations is expressed entirely thorough the
variable $a(t)$ as $a(t)=\frac{\alpha_{s}}{4\pi}$. Note that we
used the Laplace $s$-space of the splitting functions as they are
given by $\Phi_{f}(s)=\mathcal{L}[P_{qg}(x);s]$,
$\Phi_{g}(s)=\mathcal{L}[P_{gq}(x);s]$,
$\Theta_{f}(s)=\mathcal{L}[P_{qq}(x);s]$ and
$\Theta_{g}(s)=\mathcal{L}[P_{gg}(x);s]$. The splitting functions
$P_{ij}^{,}s$ are the LO and $N^{n}LO$ Altarelli- Parisi splitting
kernels at one and high loops corrections presented in
Refs.[13-14] which satisfy the following expansion
\begin{eqnarray}
P_{ij}(x,\alpha_{s}(Q^{2}))&=&\frac{\alpha_{s}}{4\pi}P_{ij}^{\rm
LO}(x)+(\frac{\alpha_{s}}{4\pi})^{2}P_{ij}^{\rm
NLO}(x)\nonumber\\
&& +(\frac{\alpha_{s}}{4\pi})^{3} P_{gg}^{\rm NNLO}(x)+.....
\end{eqnarray}
The running coupling constant $\alpha_{s}$ has the following forms
in NLO up to NNLO respectively [15]
\begin{equation}
\alpha_{s}^{\rm
NLO}=\frac{4\pi}{\beta_{0}t}[1-\frac{\beta_{1}{\ln}t}{\beta_{0}^{2}t}],
\end{equation}
and
\begin{eqnarray}
\alpha_{s}^{\rm
NNLO}&=&\frac{4\pi}{\beta_{0}t}[1-\frac{\beta_{1}{\ln}t}{\beta_{0}^{2}t}+\frac{1}{(\beta_{0}t)^{2}}
[(\frac{\beta_{1}}{\beta_{0}})^{2}\nonumber\\
&&(\ln^{2}t-{\ln}t+1)+\frac{\beta_{2}}{\beta_{0}}]].
\end{eqnarray}
where $\beta_{0}=\frac{1}{3}(33-2N_{f})$,
$\beta_{1}=102-\frac{38}{3}N_{f}$ and
$\beta_{2}=\frac{2857}{6}-\frac{6673}{18}N_{f}+\frac{325}{54}N_{f}^{2}$.
The variable $t$ is defined as
$t={\ln}(\frac{Q^{2}}{\Lambda^{2}})$ and $\Lambda$ is the QCD
cut- off parameter at each heavy quark mass threshold as we take the $N_{f}=4$ for $m_{c}^{2}<\mu^{2}<m^{2}_{b}$.\\
When referring to decoupling between differential equations (i.e.
Eqs.1-2), the Laplace transform method exhibits two second-order
differential evolution equation
 for singlet and gluon distribution function separately.
 In $s$-space these
equations have the following forms:
\begin{widetext}
 \begin{eqnarray}
\frac{\partial^{2} g(s,t)}{\partial
t^{2}}&=&[-(\sum_{n=1}a^{n}(t)\Theta_{g}^{(n)}(s))\frac{\partial}{\partial
t}(\frac{1}{(\sum_{n=1}a^{n}(t)\Theta_{g}^{(n)}(s))})+\sum_{n=1}a^{n}(t)(\Phi_{f}^{(n)}(s)+\Phi_{g}^{(n)}(s))]\frac{\partial
g(s,t)}{\partial t}\nonumber\\
&&+[\sum_{n=1}a^{n}(t)\Theta_{g}^{(n)}(s) \frac{\partial}{\partial
t}(\frac{\sum_{n=1}a^{n}(t)\Phi_{g}^{(n)}(s)}{\sum_{n=1}a^{n}(t)\Theta_{g}^{(n)}(s)})-
\sum_{n=1}a^{n}(t)\Phi_{f}^{(n)}(s)\sum_{n=1}a^{n}(t)\Phi_{g}^{(n)}(s)\nonumber\\
&&+\sum_{n=1}a^{n}(t)\Theta_{g}^{(n)}(s)\sum_{n=1}a^{n}(t)\Theta_{f}^{(n)}(s)]g(s,t),\\
\frac{\partial^{2} f_{s}(s,t)}{\partial
t^{2}}&=&[-(\sum_{n=1}a^{n}(t)\Theta_{f}^{(n)}(s))\frac{\partial}{\partial
t}(\frac{1}{(\sum_{n=1}a^{n}(t)\Theta_{f}^{(n)}(s))})+\sum_{n=1}a^{n}(t)(\Phi_{f}^{(n)}(s)+\Phi_{g}^{(n)}(s))]\frac{\partial
f_{s}(s,t)}{\partial t}\nonumber\\
&&+[\sum_{n=1}a^{n}(t)\Theta_{f}^{(n)}(s) \frac{\partial}{\partial
t}(\frac{\sum_{n=1}a^{n}(t)\Phi_{f}^{(n)}(s)}{\sum_{n=1}a^{n}(t)\Theta_{f}^{(n)}(s)})-
\sum_{n=1}a^{n}(t)\Phi_{f}^{(n)}(s)\sum_{n=1}a^{n}(t)\Phi_{g}^{(n)}(s)\nonumber\\
&&+\sum_{n=1}a^{n}(t)\Theta_{g}^{(n)}(s)\sum_{n=1}a^{n}(t)\Theta_{f}^{(n)}(s)]f_{s}(s,t).
 \end{eqnarray}
 \end{widetext}
In order to find solutions for distribution functions (i.e.
$G(x,t)$ and $F_{2}^{s}(x,t)$) we consider the inverse Laplace
transform of splitting functions in $s$-space. One can determine
these functions for the decoupled second order differential
equations (i.e. Eqs.6 and 7) in terms of the initial
distributions. Solving these equations in $x$-space and taking all
the above considerations into account, we find
 \begin{eqnarray}
\frac{\partial^{2} G(x,t)}{\partial
t^{2}}&=&[Eg1]{\otimes}\frac{\partial G(x,t)}{\partial
t}+[Eg2]{\otimes}G(x,t),\\
\frac{\partial^{2} F_{2}^{s}(x,t)}{\partial
t^{2}}&=&[Es1]{\otimes}\frac{\partial F_{2}^{s}(x,t)}{\partial
t}+[Es2]{\otimes}F_{2}^{s}(x,t).
 \end{eqnarray}
The inverse Laplace transform of brackets in Eqs.(6) and (7) are
defined as kernels $Egi$ and $Esi$ ($i$=1 and 2) respectively. We
firstly refer to the $\nu$-space as
$Efi(\nu,t){\equiv}\mathcal{L}^{-1}[[....];\nu]$ ($f=g, s$) then
define $\nu{\equiv}\ln(1/x)$. Therefore the decoupled solutions of
the DGLAP evolution equations with respect to $x$ and
$t(\mathrm{or}~Q^{2})$ variables are obtained. These results are
completely general and give the gluon and singlet distribution
functions at leading order up to high-order corrections.\\
\section{3. Decoupling DGLAP+GLRMQ}
When $x$ is small, annihilation comes into play as gluon density
increases in a phase space sell $\Delta{\ln\frac{1}{x}}\Delta{\ln
Q^{2}}$. In a phase space, the number of partons increases through
gluon splitting and decreases through gluon recombination. This
behavior for the singlet and gluon distribution functions has been
derived by GLR-MQ [5] as the GLRMQ evolution equation in terms of
the gluon distribution function can be expressed as
\begin{eqnarray}
\frac{{\partial}G(x,Q^{2})}{{\partial}{\ln}Q^{2}}&=&\frac{{\partial}G(x,Q^{2})}{{\partial}{\ln}Q^{2}}|_{DGLAP}\nonumber\\
&&-\frac{81}{16}\frac{\alpha^{2}_{s}}{R^{2}Q^{2}}\int^{1}_{x}\frac{dy}{y}[G(y,Q^{2})]^{2},\\
\frac{{\partial}F_{2}^{s}(x,Q^{2})}{{\partial}lnQ^{2}}&=&\frac{{\partial}F_{2}^{s}(x,Q^{2})}{{\partial}lnQ^{2}}|_{DGLAP}\nonumber\\
&&-\frac{27\alpha_{s}^{2}}{160R^{2}Q^{2}}[xg(x,Q^{2})]^{2}+G_{HT}(x,Q^{2}).\nonumber\\
\end{eqnarray}
The first terms in the above equations are the usual linear DGLAP
terms, and the second terms in Eqs.(10) and (11) control the
strong growth by the linear terms. The higher dimensional gluon
term $G_{TH}$, where $HT$ denotes a further term notified by
Mueller and Qiu [5], is assumed to be zero. Also the quark gluon
emission diagrams, due to their little importance in the gluon
rich, neglected. Indeed the interaction of the second gluon, when
it is situated just behind another one, is not need to count if we
have taken into account the interaction of the first one. These
nonlinear corrections can be control by parameter
$\kappa=(\frac{3\pi^{2}\alpha_{s}}{2Q^{2}}{\times}\frac{G(x,Q^{2})}{\pi
R^{2} })$. It means that in a large kinematic region $\kappa \geq
1 $ for high gluon density QCD (hdQCD), we expect that the
nonlinear corrections should be large and important for a
description of the LHeC data.\\
Defining the Laplace transforms of the nonlinear terms in Eqs.(10)
and (11) and using the convolution factors in this transformation
in $s$-space as we have
\begin{eqnarray}
\frac{\partial f_{s}(s,t)}{\partial
t}&=&[Eq.1]-T(R,Q^{2})g(s,t)^{2} ,\\
 \frac{\partial
g(s,t)}{\partial t}&=&[Eq.2]-\frac{K(R,Q^{2})}{s}g(s,t)^{2}.
 \end{eqnarray}
where $T(R,Q^{2})=\frac{27\alpha_{s}^{2}}{160R^{2}Q^{2}}$ and
$K(R,Q^{2})=\frac{81\alpha_{s}^{2}}{16R^{2}Q^{2}}$. The value of
$R$ is the correlation radius between two interacting gluons. The
correlation radius is the order of the proton radius
$(R\simeq5\hspace{0.1cm} GeV^{-1})$ if gluons are distributed
through the whole of proton, or much smaller
$(R\simeq2\hspace{0.1cm} GeV^{-1})$ if gluons are concentrated in
hot- spot within the proton.\\
One can rewrite the nonlinear terms into
 the $s$-space  as  the
nonlinear equations (i.e., Eqs.(12) and (13) )decoupled into the
second order differential equations by the following forms
\begin{widetext}
 \begin{eqnarray}
\frac{\partial^{2} g(s,t)}{\partial
t^{2}}&=&[Eq.6]-[(\sum_{n=1}a^{n}(t)\Theta_{g}^{(n)}(s))\frac{\partial}{\partial
t}(\frac{K(R,t)}{(\sum_{n=1}a^{n}(t)\Theta_{g}^{(n)}(s))})-(\sum_{n=1}a^{n}(t)\Phi_{f}^{(n)}(s))K(R,t)]\frac{g^{2}(s,t)}{s}\nonumber\\
&&-[(\sum_{n=1}a^{n}(t)\Theta_{g}^{(n)}(s))T(R,t)]g^{2}(s,t)-\frac{K(R,t)}{s}\frac{\partial
g^{2}(s,t) }{\partial t},\\
 \frac{\partial^{2}
\digamma(s,t)}{\partial
t^{2}}=&=&[Eq.7]+\frac{5}{18}[-(\sum_{n=1}a^{n}(t)\Theta_{f}^{(n)}(s))\frac{\partial}{\partial
t}(\frac{T(R,t)}{(\sum_{n=1}a^{n}(t)\Theta_{f}^{(n)}(s))})+(\sum_{n=1}a^{n}(t)\Phi_{g}^{(n)}(s))T(R,t)]g^{2}(s,t)\nonumber\\
&&-[(\sum_{n=1}a^{n}(t)\Theta_{f}^{(n)}(s))K(R,t)]\frac{g^{2}(s,t)}{s}-K(R,t)\frac{\partial
g^{2}(s,t) }{\partial t}.
 \end{eqnarray}
 \end{widetext}
The one- and two-loop splitting functions (LO and NLO )in
$s$-space for the parton distributions have been known in Refs.[9]
and [12] respectively. In $x$-space, authors in Refs.[13] and [14]
presented the exact results as well as compact parameterizations
for the splitting functions  at small $x$. The results and various
aspects of
those have been discussed in Ref.[13].\\
Our prediction at small-$x$ is a transition from the linear to the
nonlinear regime. In this region we expect to observe the
saturation of the growth of the gluon and singlet densities. At
high order corrections the ladder gluons are coupled together. We
observe that a transition to the triple-Pomeron vertex is occurred
in the leading $\ln k^{2}$ approximation than a simple Pomeron.
Indeed the nonlinear evolution equation is a crude approximation
in the double logarithmic limit. The high energy
factorization formula is an approach beyond this limit.\\
The correct degrees of freedom are given by $q\overline{q}$
colorless dipoles for high energy $\gamma^{*}p$ scattering.
 This behavior is related to the dipole cross section
$\sigma (x,r)$ where  $r$ is the $q\overline{q}$ dipole transverse
separation. The dipole cross section is given by [20]
\begin{eqnarray}
\sigma_{q\overline{q}}=\sigma_{0}\mathcal{N}(x,\mathbf{r}),
\end{eqnarray}
where $\sigma_{0}$ is a constant which assures unitarity of the
proton structure function. Here
$\mathcal{N}(x,\mathbf{r})=1-\exp(-\frac{r^{2}}{4R_{0}^{2}})$ is
the dipole scattering amplitude and $R_{0}$ denotes the saturation
radius which $R_{0}^{2}=\frac{1}{GeV^{2}}(x/x_{0})^{\lambda}$. The
parameters of the model were found  from a fit to the data with
$x<10^{-2}$ as $\sigma_{0}=23~ \mathrm{mb}$, $\lambda{\simeq}0.3$
and $x_{0}=3.10^{-4}$.\\
The virtual photon-proton cross sections $\sigma_{T}$ and
$\sigma_{L}$ are given by
\begin{eqnarray}
\sigma_{T,L}=\int d^{2}\mathbf{r}dz
|\Psi_{T,L}(\mathbf{r},z,Q^{2})|^{2}\sigma_{q\overline{q}},
\end{eqnarray}
where $\Psi_{T}$ and $\Psi_{L}$ are the light-cone wave function
for the transverse and longitudinal polarized virtual photons.
These cross sections are related to the proton structure function
by the following form
\begin{eqnarray}
F_{2}(x,Q^{2})=\frac{Q^{2}}{4\pi^{2}
\alpha_{em}}(\sigma_{T}+\sigma_{L}).
\end{eqnarray}
\section{4. The behavior of the distribution functions}

In order to show our results we computed the singlet and gluon
distribution functions into the decoupled second order evolution
equations. The initial conditions starts at $Q_{0}^{2}=1~ GeV^{2}$
with $\alpha_{s}(1~ GeV^{2} )=0.49128$ and
$\alpha_{s}(M_{z}^{2})=0.12018$ at NLO analysis [16]. The power
law behavior of the distribution functions is considered  in the
decoupled DGLAP evolution equations. In principle if we take
$F_{2}{\sim}x^{-\lambda_{s}}$ and $G{\sim}x^{-\lambda_{g}}$ then
we might expect to determine $\lambda_{s}$ and $\lambda_{g}$. The
behavior of exponent $\lambda_{s}$ at fixed $Q^{2}$ should be
determined from Eq.9 by the proton structure function extracted
from
Ref.[17].\\
As can be seen in Fig.1, these derivatives are independent of $x$
when compared with H1 2001 data [18] within the experimental
accuracy. In this figure we show $\lambda_{s}$ calculated  as a
function of $x$ for $5\leq Q^{2}\leq 250~ GeV^{2}$ from H1 2013
data [17]. These result are consistent with other experimental
data [19]. In order to test the validity of our obtained
exponents, we introduce the averaged value  $\lambda_{i},~(i=s~
\mathrm{and}~g)$ (denoted in the following by $<...>$) in this
research. In Fig.2 we show the NLO results for the proton
structure function as a function of $<\lambda_{s}>$ as compared
with H1 collaboration data [17].\\
 It is tempting, however, to
explore the possibility of obtaining approximate behavior of the
exponents  in the restricted color dipole model point (i.e.,
$x_{0}$). In Figs.3-4 the approximate behavior of the exponents
are shown for $|\Delta{\lambda}_{i}|=\lambda_{i}-<\lambda_{i}>$.
These data are obtained with respect to the decoupled DGLAP
evolution equations in accordance with H1 data. We observed that
$|\Delta{\lambda}|$ behaviors have a derivative at around a saddle
point. Our fit to these results show that a minimal point (i.e.,
$x_{cut}$) is corresponding to the derivatives. As seen in these
figures the saddle point is almost constant in a wide range of
$Q^{2}$ values. This point
approximately has the same value as observed for the color dipole model point (i.e. $x_{cut}\simeq x_{0}$).\\
 Therefore we expect to have the intercepts for
$x<x_{0}$ and $x>x_{0}$ where the averaged value is taken as a
constant factor throughout the calculation.
 The results of
the average to data are collected in Tables I and II for $x<x_{0}$
and $x>x_{0}$ respectively. In these tables a transition from the
soft pomeron to the hard pomeron intercept for distribution
functions observed as $Q^{2}$ increase. We observed that the
low-$x$ behavior of the both gluon and sea quarks is controlled by
the pomeron intercept.\\
In Fig.5 we show $<\lambda_{i}>$ obtained for singlet structure
function and gluon distribution function at $x<x_{0}~ \mathrm{and}
~x>x_{0}$ and compared the singlet averaged value of
$<\lambda_{s}>$ with $\lambda_{\mathrm{phn}}(Q^{2})$ parameterized
in Ref.[21]. This phenomenologically exponent has been derived
(Ref.[21]) for calculating the evolution of singlet density for
combined HERA $e^{+}p$ DIS data [22] within the saturation model.
With this method the low-$x$ behavior of the $F_{2}$ structure
function has been shown that
$F_{2}(x,Q^{2}){\sim}x^{-\lambda_{phn}(Q^{2})}$, where
$\lambda_{phn}(Q^{2})$ can be parameterized as
$\lambda_{phn}(Q^{2})=0.329+0.1 \log(\frac{Q^{2}}{90})$.\\
In this figure (i.e. Fig.5) we observe that a linear behavior for
exponents $<\lambda_{s}>$ at low-$Q^{2}$ values is dominant and
this behavior is almost constant at high-$Q^{2}$  values. It is
observed that the averaged exponent $<\lambda_{s}>$ shows almost
similar behavior with $x<x_{0}~ \mathrm{and} ~x>x_{0}$ but in
comparison with $\lambda_{phn}$ it shows that this behavior is
nonlinear in wide range of $Q^{2}$ values. We also added results
for the gluon exponent  at low and high $Q^{2}$ values in the same
figure. Similar remarks apply to the gluon exponent where
 the scale of the exponent has been fixed at a hard pomeron value. \\
One can see that exponents obtained for gluon distribution
function from the decoupled evolution equations are larger than
the singlet exponent, that is $\lambda_{g}>\lambda_{s}$. Indeed
the steep behavior of the gluon generates a similar steep behavior
of singlet at small-$x$ at NLO analysis where
$\lambda_{s}=\lambda_{g}-\epsilon$. Furthermore, the exact
solution of the decoupled evolution equations predicts that
$\lambda_{s}$ and $\lambda_{g}$ are separated free parameters.
This shows that the differences between these exponents are
consistent with pQCD. In Fig.6 we compared $\Delta{\lambda}$
obtained using an averaged value between singlet and gluon
exponents at $x<x_{0}~ \mathrm{and} ~x>x_{0}$. This
$\Delta{\lambda}$ shows that a scaling behavior property is
exhibited for $Q^{2}>5~GeV^{2}$. Note that this scaling for
singlet and gluon exponents in Fig.5 is clear at moderate and high
$Q^{2}$ values.\\
In Figures 7 and 8 we present the results for the proton structure
function and the gluon distribution function at NLO analysis using
the gluon and singlet exponents for $x<x_{0}~ \mathrm{and}
~x>x_{0}$ respectively. We compare our results with those obtained
by GJR parameterizations [23] and H1 2013 experimental data [17].
The kinematical ranges for H1 2013 data are $10^{-7}<x<10^{-1}$
and $5 \leq Q^{2} \leq 250~GeV^{2}$. These results in Fig.7
compared with H1 data  and GJR parameterizations in
 a wide range of $x$ values  based on averaged exponents for $x<x_{0}~ \mathrm{and}
~x>x_{0}$ . Also, in Fig.8,
 the gluon distribution behavior is compared  with GJR
 parameterization with respect to the averaged exponent. These
 results indicate that our calculations, based on the averaged
 exponent for $x>x_{0}$ have  of the same form behavior as the
 predicted by other results. But for $x<x_{0}$
the extrapolation to small $x$ has very large uncertainties when
compared with GJR parameterization especially at large $Q^{2}$
values. This is because the averaged exponent behavior defined by
our method (in Table I) is expected to hold in the small $x$
limit. However, due to the existence of absorptive corrections,
this is not true pomeron intercept, but rather an effective one is
necessary in this region [24]. The connection between the averaged
value of exponent
   with effective exponent and color dipole model size is
 given in next sections.\\
\section{5. Effective Exponents}
Perturbative QCD predicts a strong power-law rise of the gluon and
singlet  distribution  at low $x$. This behavior coming from
resummation of large powers of $\alpha_{s} \ln1/x$ where its
achieved by the use of the $k_{T}$ factorization formalism. The
small-$x$ resummation requires an all-order class of subleading
corrections in order to lead to stable results. Authors in
Ref.[25] discuss a framework to perform small-$x$ resummation for
both parton evolution and partonic coefficient functions. In this
paper a detailed analysis has been performed in order to find
resummation of the DGLAP and BFKL evolution kernels at NNLO
approximation.\\
In the leading log approximation this behavior leads to the
unintegrated gluon distribution rising as a power of $x$. Which
this result is given in terms of the BFKL evolution equation as
$f(x,k_{T}^{2})\sim x^{-\lambda}$ [20]. The function
$f(x,k_{T}^{2})$ is the unintergrated gluon distribution and
related to the gluon distribution from the DGLAP evolution
equation by integration over the transform momentum as
$G(x,Q^{2})=\int^{Q^{2}}\frac{dk_{T}^{2}}{k_{T}^{2}}f(x,k_{T}^{2})$.
One finds $\lambda = 0.437$ where it is the so-called hard-Pomeron
exponent. As an alternative to comparing with the experimental
data an effective exponent $\lambda\simeq 0.3$ exhibited in Refs.
[20-21]. The strong rise into the $k_{T}$ factorization formula is
also true for the singlet structure function. The BFKL Pomeron
does not depend on $Q^{2}$, however the effective Pomeron is
$Q^{2}$-dependent when structure functions fitted to the
experimental data at low values of $x$. This behavior can violate
unitarity, so it has to be tamed by screening effects.\\
Indeed the nonlinear terms reduce the growth of singlet and gluon
distributions at low-$x$. The transition point for gluon
saturation is given by the saturation scale
$Q_{s}^{2}=Q_{0}^{2}(\frac{x}{x_{0}})^{-\lambda}$ which is the
$x$-dependent and this is an intrinsic characteristic of a dense
gluon system. Here, we take into account the effects of kinematics
which these are shown results a shift from the pomeron exponent to
the effective exponent. This is related to fact that at low $x$ we
needed to produce the color dipole model in the argument of the
gluon distribution as it can be computed from the $k_{T}$
factorization formula [26]. We note that the nonlinear effects are
small for $Q^{2}>Q_{s}^{2}$, but very strong for $Q^{2}<Q_{s}^{2}$
where leading to the saturation of the scattering amplitude. These
two regions separated by the saturation line, $Q^{2}=Q^{2}_{s}$.
When $x\rightarrow x_{0}$, we have $Q^{2}=Q^{2}_{s}\rightarrow 1$
and this line is independent of $\lambda$-exponent. With respect
to the averaged values of $\lambda_{s}$ for $x<x_{0}~ \mathrm{and}
~x>x_{0}$, the relation $Q^{2}>Q_{s}^{2}$ is  satisfied always
when we used the HERA kinematical region for $5 \leq Q^{2} \leq
250~GeV^{2}$ and $10^{-6} \leq x \leq 10^{-2}$. However we do not
observe the
nonlinear behavior by these exponents.\\
In Tables III and IV, we show some of parameters determined for
the saturation behavior of the singlet structure function. These
saturation parameters are obtained at $Q^{2}=5~GeV^{2}$ and
$250~GeV^{2}$ respectively. We note that the averaged exponent
$<\lambda_{s}>$ is taken into account from Tables I and II. In
Tables III and IV, the saturation scale is defined in the form
$R_{0}^{2}(x)=\frac{1}{Q_{0}^{2}}(\frac{x}{x_{0}})^{\lambda}$
where $Q_{0}^{2}=1~GeV^{2}$. Indeed $R_{0}(x)$ decreases when
$x\rightarrow 0$. Here the dimensionless variable $\tau$ is taken
with a simplest form $\tau=Q^{2}R_{0}^{2}(x)$ throughout
the method [20].\\
The saturation radius is defined by $r_{s}=2R_{0}(x)$ because its
decrease with decreasing $x$. Indeed if $r<r_{s}(x)$ the dipole
cross section increase as $x$ decreases. We note that the variable
$r$ denotes the separation between the quark and antiquark in
color dipole model. This transverse dimension of the
$q\overline{q}$ pair is small when the condition $r<\frac{1}{Q}$
is fulfilled and large when $r>\frac{1}{Q}$.
 In Fig.9 we analyse  the ratio of the dipole cross
section, $\sigma/\sigma_{0}$, for different dimensionless variable
$\tau$. As expected the saturation lies in the small-$r$ region
when compared the region $x<x_{0}$ with $x>x_{0}$.\\
 To better illustrate nonlinear effects at
$Q^{2}<Q^{2}_{s}$, we average over the exponents for $5\leq
Q^{2}\leq 250~GeV^{2}$ presented in Tables I and II.  As optimal
values of parameters $\lambda_{s}$ and $\lambda_{g}$ we take
$\lambda_{ave}$ over the two different regimes as we have
\begin{eqnarray}
<\lambda_{s}>=0.295~~ \mathrm{for} ~~x<x_{0}~~~ \mathrm{and}~~~
 0.358~~ \mathrm{for} ~~x>x_{0},\nonumber
\end{eqnarray}
\begin{eqnarray}
<\lambda_{g}>=0.362~~ \mathrm{for}
~~x<x_{0}~~~\mathrm{and}~~~0.485~~ \mathrm{for}
~~x>x_{0}.\nonumber
\end{eqnarray}
Therefore the averaged value to all exponents has the effective
constraint as we obtained the following effective exponents and
taking all into account.
\begin{eqnarray}
\lambda^{eff}_{s}=0.327 ~ \mathrm{and} ~ \lambda^{eff}_{g}=0.424.
\end{eqnarray}
One can see that the averaged exponents obtained for singlet and
gluon distributions are closer to those defined by color dipole
model and hard-pomeron exponents. In Ref.[21], authors have shown
that the  inclusive DIS value of singlet exponent is defined as
$\lambda_{inc}=0.298 \pm 0.011$, for combined HERA DIS data where
errors are purely statistical. This parameter is defined to be
$0.29$ in GBW model [20] and it is an effective intercept in BFKL
kernel. Also $\lambda_{g}$ is comparable with the so-called hard
Pomeron intercept [27].\\
 Now the saturation condition
$Q^{2}<Q_{s}^{2}$ is visible for low $x$ values at low and
moderate $Q^{2}$ values. As illustrated in Fig.10 the transition
occurs for decreasing transverse sizes at very low $x$ values. We
observe a continuous behavior of  the  dipole cross section
saturation towards small-$r$. The saturation form of the ratio
$\sigma/\sigma_{0}$ is modified in comparison with Fig.9 based on
the effective exponents. Therefore these exponents (i.e. Eq.19)
guarantees consistency low $x-Q^{2}$
behavior with saturation effects. The proton structure function behavior with respect to this effective exponent  will study in next section.\\
\section{6. The behavior of the structure function}
In order to show the saturation effects at low values of $x$, we
computed the proton structure function by the effective exponent
obtained in Eq.19. In Fig.11 the $F_{2}$ structure function is
plotted as a function of $x$ in bins of $Q^{2}$ at NLO analysis.
We compare our results with H1 experimental data [17-19,22]. The
agreement between the experimental data and our calculation at
moderate-$Q^{2}$ values is good, as the exponent value $\lambda
\sim 0.33$ is served for inclusive DIS and other methods. At low
and high-$Q^{2}$ values we observed an overall shift between the
H1 data and the predictions. This behavior can be resolved with an
adjustment of lower and higher  exponents than one obtained for
decoupled distributions respectively.\\
 Notice that this behavior for the effective exponent
is closer than to the linear behavior with $\log Q^{2}$ as given
in Ref.21. Indeed this behavior depends on fixed effective
exponent which we constrain analyzing for $Q_{s}$. In Fig.11 this
prediction obtained at very low-$x$ values as the exponent is
fixed in accordance with an effective intercept. It is tempting,
however, to explore the possibility of obtaining an effective
exponent dependent on $Q^{2}$ in the restricted domain of all
$Q^{2}$ values at least. In Fig.12, $\lambda (Q^{2})$ depicted and
compared with the linear dependence on $\log Q^{2}$ in Ref.[21].
To better illustrate our calculations at all $Q^{2}$ values, we
used therefore effective exponent in the form of $\lambda(Q^{2})$
in Fig.13. This figure indicate that the obtained results from
present analysis based on $\lambda(Q^{2})$ are in good agreements
with H1 data for the proton structure function. These results have
been presented as a function of $x$ (both for large and small $x$
) at $5 \leq Q^{2} \leq 250~GeV^{2}$.
 In transition to
the saturation model for low $Q^{2}$, one should take into account
the quark mass, when we replace
$x{\rightarrow}(1+\frac{4m_{c}^{2}}{Q^{2}})x$ in decoupled
evolution equations [28]. Indeed one can modify
$\tau=\frac{Q^{2}}{Q^{2}_{0}}(\frac{x}{x_{0}})^{\lambda}$ to take
into account $m_{c}\simeq 1.3~GeV^{2}$ and introduce scaling
$\tau_{c}=(1+\frac{4m_{c}^{2}}{Q^{2}})\frac{Q^{2}}{Q^{2}_{0}}(\frac{x}{x_{0}})^{\lambda}$
[29]. However the transition to the low and high $Q^{2}$ values is
related to the regions that $\lambda_{s}<\lambda^{eff}_{s}(\simeq
0.33)$ and $\lambda_{s}>\lambda^{eff}_{s}(\simeq 0.33)$
respectively, as we observed that the effective exponent is not a
linear function into $\log Q^{2}$. It has the nonlinear behavior
with respect to the
low and high $Q^{2}$ values.\\
\section{7. Conclusion}
We presented the high-order decoupled  analytical evolution
equations for distribution functions, arising from the coupled
DGLAP evolution equations. The Laplace transform technique used
for the decoupled proton structure function and gluon distribution
function evolution equations. Two homogeneous second-order
differential evolution equations are obtained and extended to the
nonlinear behavior at low-$x$ region. The next-to-leading -order
analysis compared with H1 data with the averaged value of exponent
$\lambda_{s}$. The averaged value of exponent $\lambda_{s}$ has
different behavior when the color dipole model is considered
around the $x_{0}$ value. The different between the gluon and
singlet exponents considered at low-$x$ values and shown that they
have the nonlinear behavior into $\log Q^{2}$. The dipole cross
section considered for nonlinear behavior at $Q^{2}<Q^{2}_{s}$, as
this behavior is very important for shown that the effective
exponent with exact value is necessary for saturation effect. This
effective exponent is basically the value of $\simeq 0.33$ as
reported in the literature. The proton structure function
determined and compared with respect to this effective exponent
for $10^{-7}\leq x \leq 10^{-2}$ and $5 \leq
Q^{2} \leq 250~GeV^{2}$.\\
\newpage
\section{References}
1. V. N. Gribov and L. N. Lipatov, Sov. J. Nucl. Phys. {\bf15}
(1972) 438; G. Altarelli and G. Parisi, Nucl. Phys. B{\bf126}
(1977) 298; Y. L. Dokshitzer, Sov. Phys. JETP{\bf46} (1977)641.\\
2. M.Klein, Ann.Phys.{\bf528}(2016)138.\\
3. P.Kostka et.al., Pos DIS2013 (2013)256; L.Han et.al.,
Phys.Lett.B{\bf 771}(2017)106; L.Han et.al., Phys.Lett.B{\bf
768}(2017)241; Yao-Bei Liu, Nucl.Phys.B{\bf923}(2017)312.\\
4. E.Gotsman et.al., Nucl.Phys.B{\bf 539} (1999)535; K.J.Eskola
et.al., Nucl.Phys.B{\bf 660} (2003)211.\\
 5. L.V.Gribov, E.M.Levin
and M.G.Ryskin, Phys.Rep.\textbf{100},
 (1983)1; A.H.Mueller and
J.Qiu, Nucl.Phys.B\textbf{268} (1986)427.\\
6. B.Rezaei and G.R.Boroun, Phys.Letts.B{\bf692}(2010)247;
G.R.Boroun, Eur.Phys.J.A{\bf42}(2009)251; G.R.Boroun,
Eur.Phys.J.A{\bf43}(2010)335.\\
7. P.Phukan et.al., arXiv:hep-ph/1705.06092; M.Lalung et.al.,
arXiv:hep-ph/1702.05291; M.Devee and J.K.sarma,
Eur.Phys.J.C{\bf74}(2014)2751; M.Devee and J.K.sarma,
Nucl.Phys.B{\bf885}(2014)571.\\
8. G.R.Boroun and S.Zarrin, Eur.Phys.J.Plus {128}(2013)119;
B.Rezaei and G.R.Boroun,  Eur. Phys. J. C{\bf73}(2013)2412;
G.R.Boroun and B.Rezaei, Eur.Phys.J.C{\bf72}(2012)2221.\\
9. Martin M.Block et al., Eur.Phys.J.C{\bf69}(2010)425; Phys.Rev.D{\bf84}(2011)094010; Phys.Rev.D{\bf88}(2013)014006.\\
10. F.Taghavi-Shahri et al., Eur.Phys.J. C{\bf71} (2011) 1590.\\
11.S.Shoeibi et al., Phys.Rev. D{\bf97} (2018) 074013.\\
12. H.Khanpour, A.Mirjalili and S.Atashbar Tehrani, Phys.Rev.C{\bf95}(2017)035201.\\
13. A. Vogt, S. Moch and J.A.M. Vermaseren,
Nucl.Phys.B{\bf691}(2004)129.\\
14. C.D. White and R.S. Thorne, Eur.Phys.J.C{\bf45} (2006)179.\\
15. B.G. Shaikhatdenov, A.V. Kotikov, V.G. Krivokhizhin, G.
Parente, Phys. Rev. D {\bf81}(2010) 034008.\\
16. A.D.Martin, et al., Eur.Phys.J.C{\bf63} (2009)189.\\
17. V. Andreev et al. (H1 Collaboration), Eur. Phys. J. C{\bf74}
(2014) 2814.\\
18. C.Adloff et al. (H1 Collaboration), Eur. Phys. J. C{\bf21}
(2001) 33.\\
19. C.Adloff et al. (H1 Collaboration), Phys.Lett.B{\bf520}
(2001) 183.\\
20. K.Golec-Biernat and M.Wuesthoff, Phys.Rev.D{\bf59} (1999) 014017; K.Golec-Biernat, Acta.Phys.Polon.B{\bf33} (2002) 2771.\\
21. M. Praszalowicz and T.Stebel, JHEP {\bf03} (2013) 090; T.Stebel, Phys. Rev. D{\bf88} (2013) 014026.\\
22. F.D.Aaron et al. (H1 and ZEUS Collaboration), JHEP{\bf1001}
(2010) 109; Eur.Phys.J.C {\bf63} (2009) 625; Eur.Phys.J.C {\bf64} (2009) 561.\\
23. M.Gluk, P.Jimenez-Delgado and E.Reya, Eur.Phys.J.C{\bf53}(2008)355.\\
24.A.Donnachie and P.V.Landshoff, Phys.Lett.B{\bf550}(2002)160; G.R.Boroun and B.Rezaei, Phys.Atom.Nucl.{\bf71}(2008)1077.\\
25. M.Bonvini et.al., Eur.Phys.J.C{\bf76}(2016)597.\\
 26. V.P.Goncalves and M.V.T.Machado, Phys.Rev.Lett.{\bf91}
(2003)
202002.\\
27. L. Motyka et al., arXiv:0809.4191v1 (2008); H.Kowalski et al., Eur.Phys.J.C{\bf77}(2017)777.\\
28. Z.Jalilian and  G.R. Boroun, Phys.Lett. B{\bf773}(2017)455.\\
29. E.Avsar and G.Gustafson, JHEP{\bf0704} (2007) 067.\\

 \begin{table}
\centering \caption{Averaged value of exponents for $Q^{2}$ values
 at $x<x_{0}$. }\label{table:table1}
\begin{minipage}{\linewidth}
\renewcommand{\thefootnote}{\thempfootnote}
\centering
\begin{tabular}{|l|c|c|} \hline\noalign{\smallskip} $Q^{2}(GeV^{2})$ & $<\lambda_{s}>$  &
$<\lambda_{g}>$  \\
\hline\noalign{\smallskip}
5& 0.135 & 0.307    \\
12& 0.262 & 0.336    \\
20& 0.295 & 0.353    \\
90& 0.337 & 0.376   \\
120& 0.342 & 0.385    \\
150& 0.344 & 0.387    \\
250& 0.351 & 0.391   \\
\hline\noalign{\smallskip}
\end{tabular}
\end{minipage}
\end{table}
 \begin{table}
\centering \caption{The same as Table I but for $x>x_{0}$.
}\label{table:table1}
\begin{minipage}{\linewidth}
\renewcommand{\thefootnote}{\thempfootnote}
\centering
\begin{tabular}{|l|c|c|} \hline\noalign{\smallskip} $Q^{2}(GeV^{2})$ & $<\lambda_{s}>$  &
$<\lambda_{g}>$  \\
\hline\noalign{\smallskip}
5& 0.135 & 0.307    \\
12& 0.312 & 0.435    \\
20& 0.356 & 0.470    \\
90& 0.419 & 0.543   \\
120& 0.411 & 0.536    \\
150& 0.425 & 0.540    \\
250& 0.445 & 0.567   \\
\hline\noalign{\smallskip}
\end{tabular}
\end{minipage}
\end{table}

\begin{figure}
\centering
\includegraphics[width=1\textwidth]{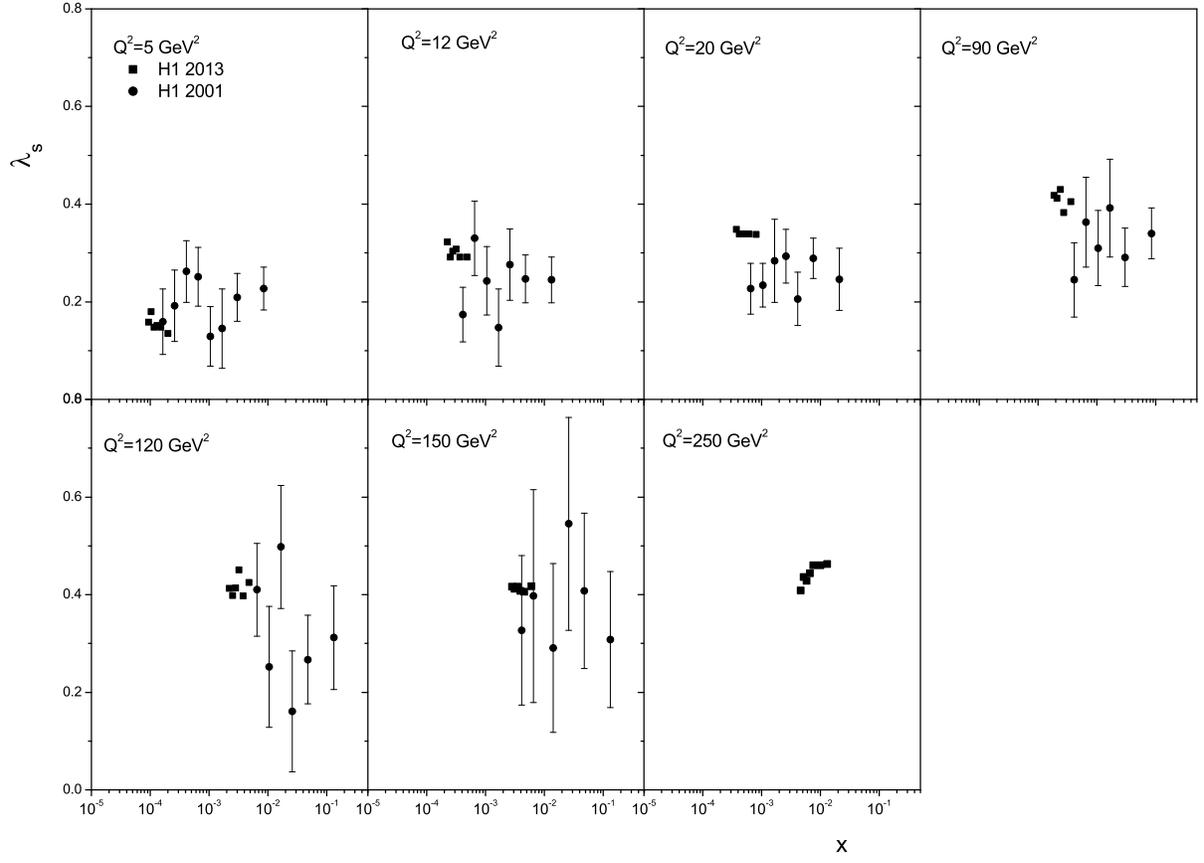}
\caption{The singlet exponent $\lambda_{s}$ from H1 2013 data [17]
compared with result H1 2001 [18-19] plotted against $x$ at $5\leq
Q^{2}\leq 250~ GeV^{2}$.} \label{Fig1}
\end{figure}
\begin{figure}
\centering
\includegraphics[width=1\textwidth]{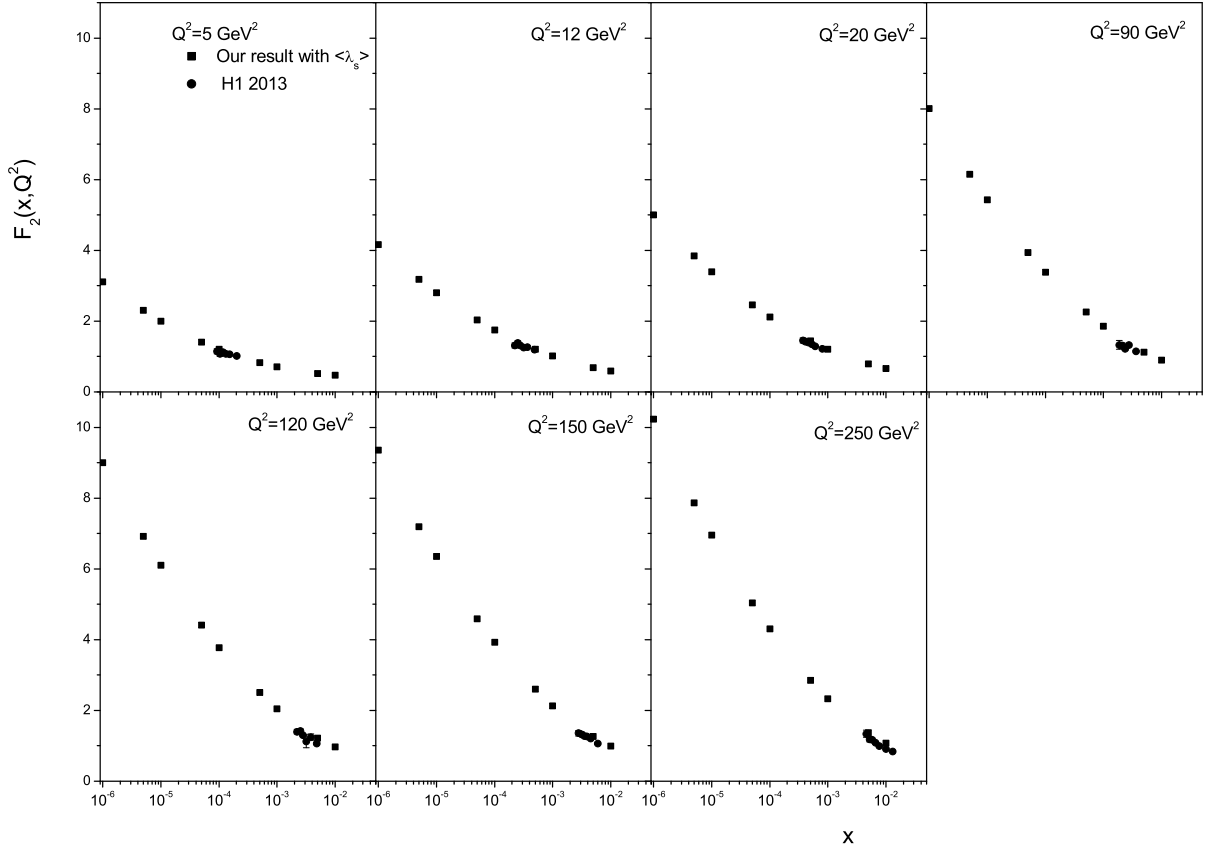}
\caption{The proton structure functions $F_{2} (x, Q^{2})$ with
respect to $<\lambda_{s}>$ compared with H1 data [17] as
accompanied with total errors.} \label{Fig2}
\end{figure}
\begin{figure}
\centering
\includegraphics[width=0.8\textwidth]{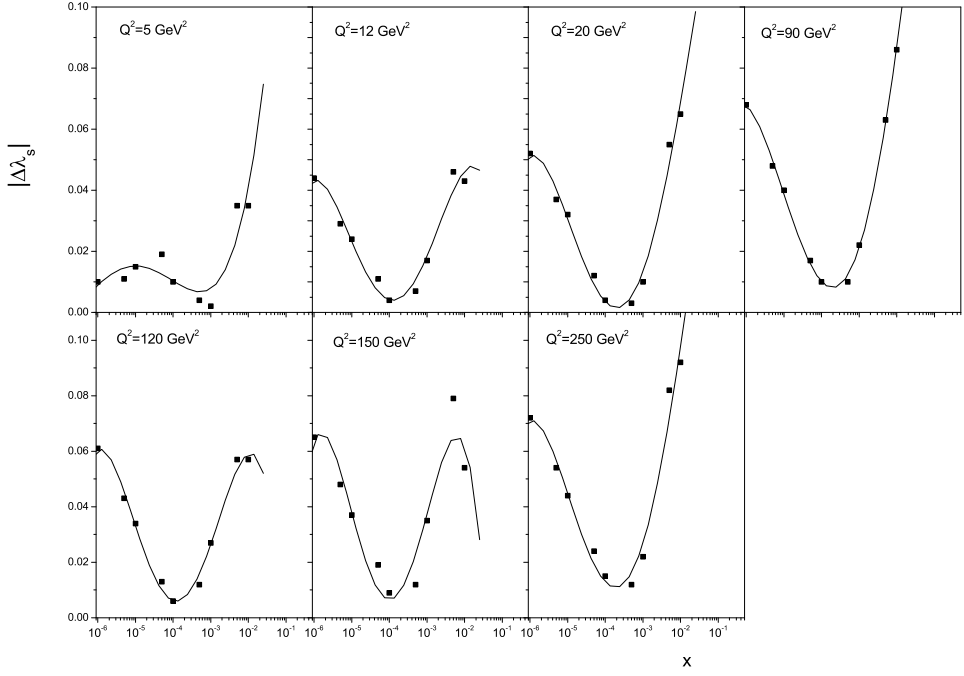}
\caption{ The singlet intercept behavior obtained for different
energies as functions of $x$. Solid curve is our fit to all data.}
\label{Fig3}
\end{figure}
\begin{figure}
\centering
\includegraphics[width=0.8\textwidth]{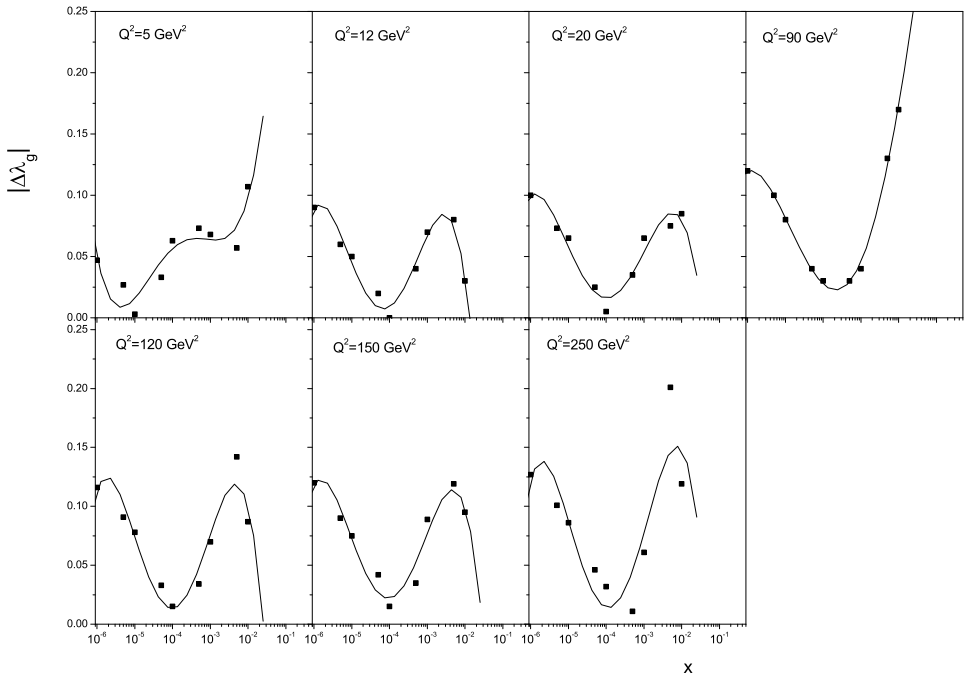}
\caption{As in Fig.3 but for the gluon exponent behavior.}
\label{Fig4}
\end{figure}
\begin{figure}
\centering
\includegraphics[width=0.6\textwidth]{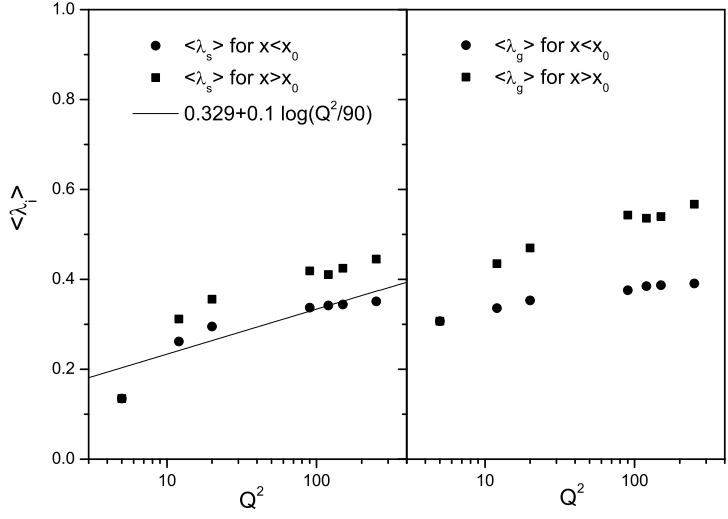}
\caption{Averaged value of singlet and gluon exponents with
respect to the critical point as functions of $Q^{2}$. The linear
fit is a effective exponent to H1 and ZEUS data in Refs.21-22.}
\label{Fig5}
\end{figure}
\begin{figure}
\centering
\includegraphics[width=0.6\textwidth]{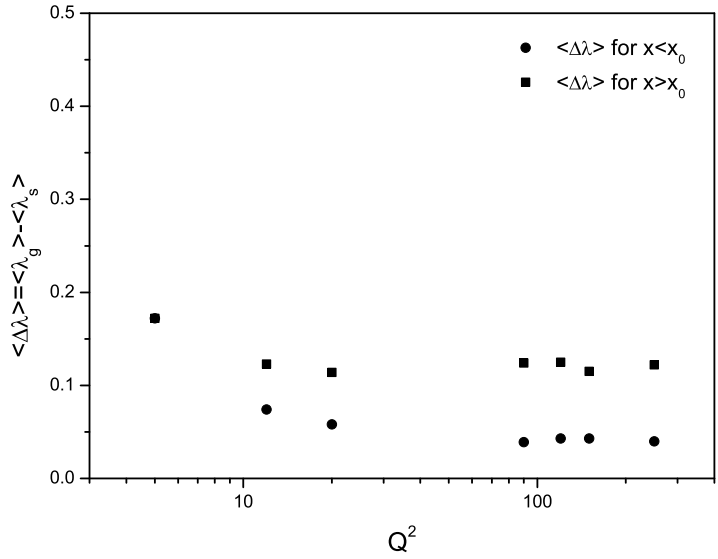}
\caption{Different of singlet and gluon exponents as functions of
$Q^{2}$.} \label{Fig6}
\end{figure}
\begin{figure}
\centering
\includegraphics[width=1\textwidth]{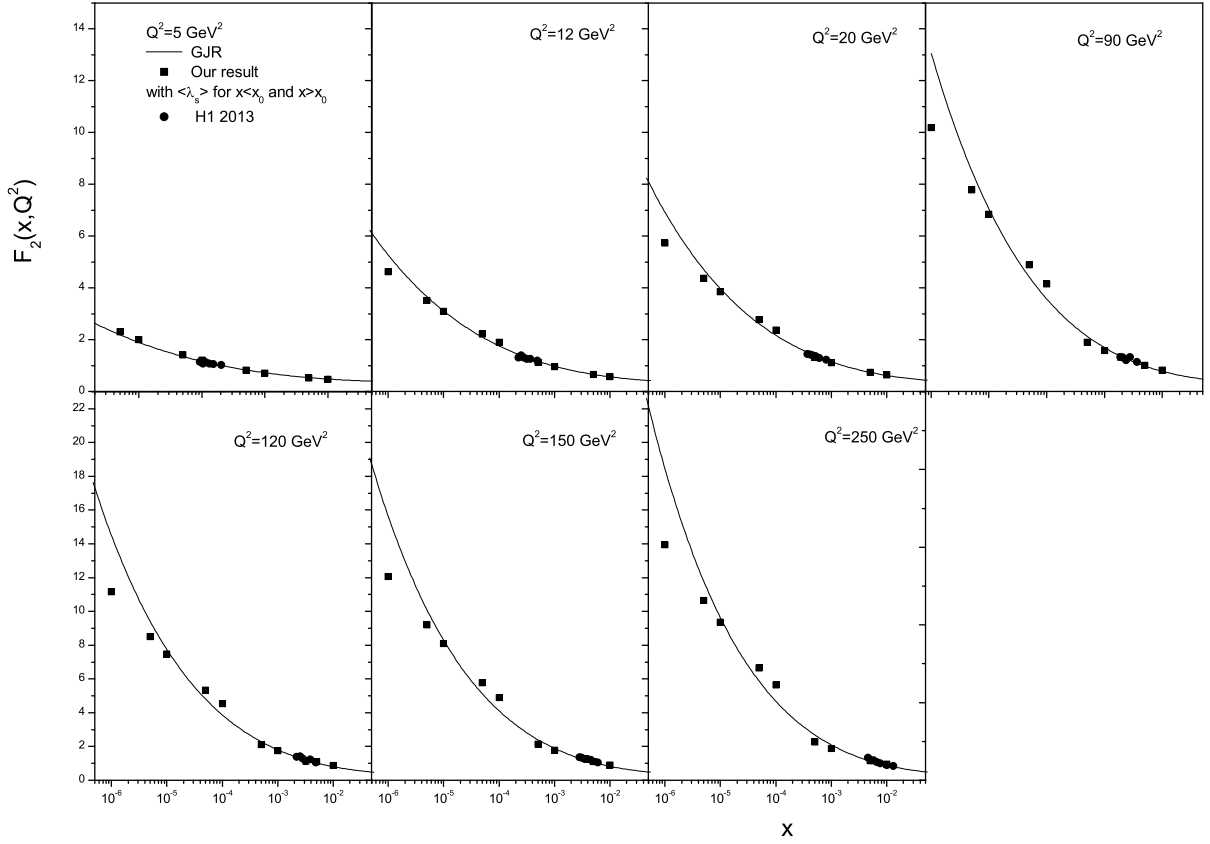}
\caption{The determined values of the structure function
$F_{2}(x,Q^{2})$ plotted as functions of $x$ with respect to the
averaged exponent $<\lambda_{s}>$ for $x<x_{0}$ and $x>x_{0}$
compared with H1 data [17] and GJR parameterization [23].}
\label{Fig7}
\end{figure}
\begin{figure}
\centering
\includegraphics[width=1\textwidth]{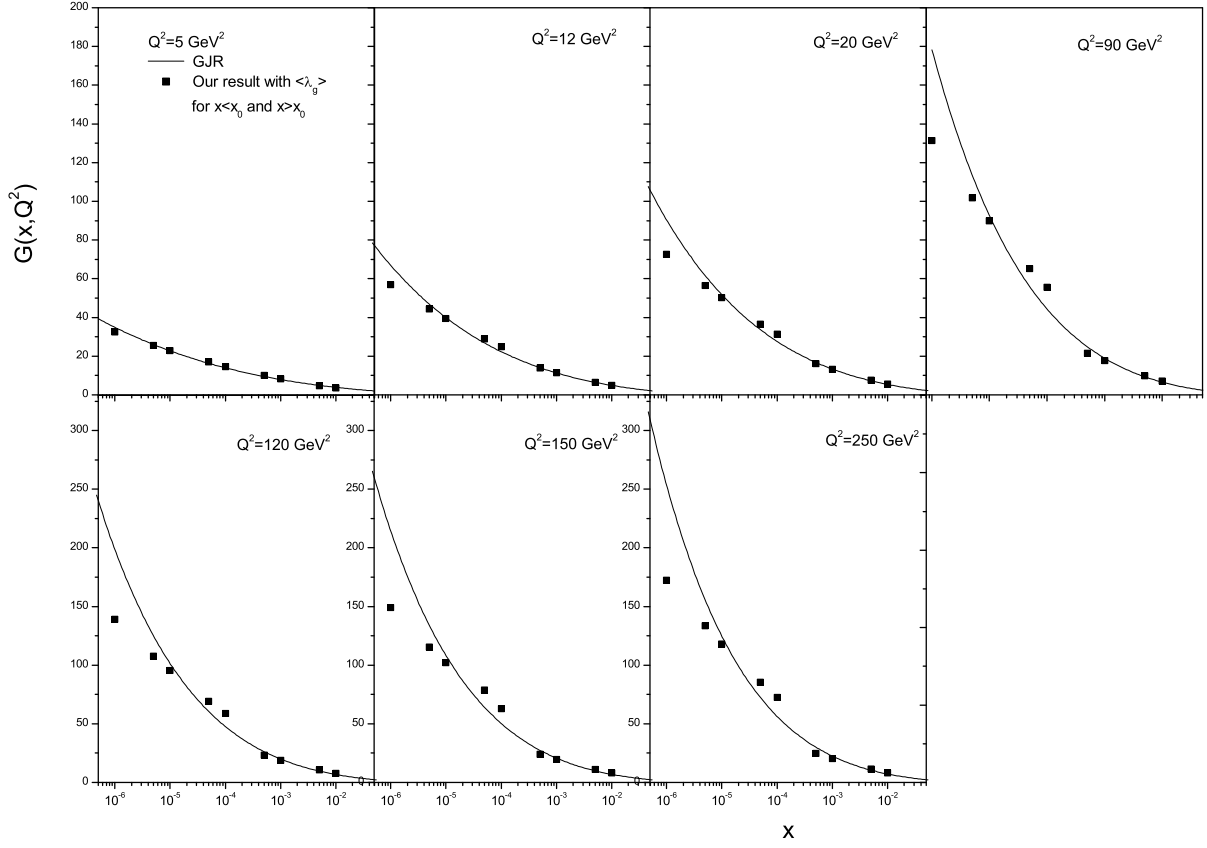}
\caption{The gluon distribution against $x$ in a wide region of
$Q^{2}$ values with respect to the averaged exponent
$<\lambda_{g}>$ for $x<x_{0}$ and $x>x_{0}$ compared with GJR
parameterization [23].} \label{Fig8}
\end{figure}
\begin{figure}
\centering
\includegraphics[width=0.6\textwidth]{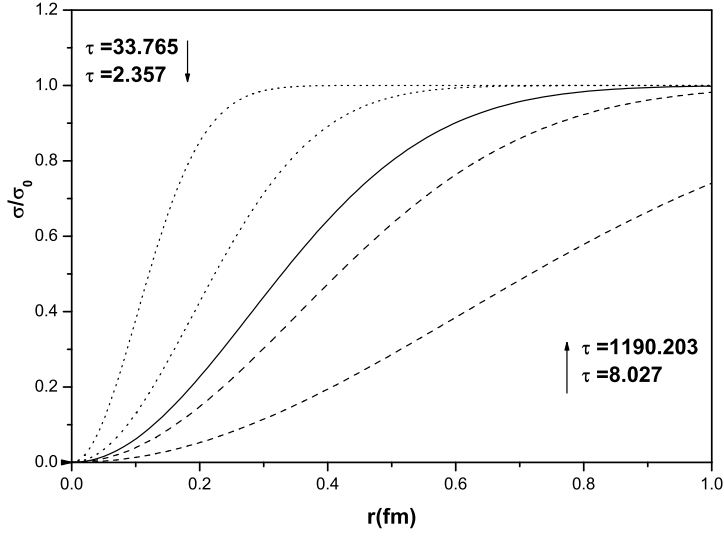}
\caption{The ratio of the dipole cross section with respect to the
averaged exponent $<\lambda_{s}>$ for $x<x_{0}~ \mathrm{and}
~x>x_{0}$ at different values of $\tau$.} \label{Fig9}
\end{figure}
\begin{figure}
\centering
\includegraphics[width=0.6\textwidth]{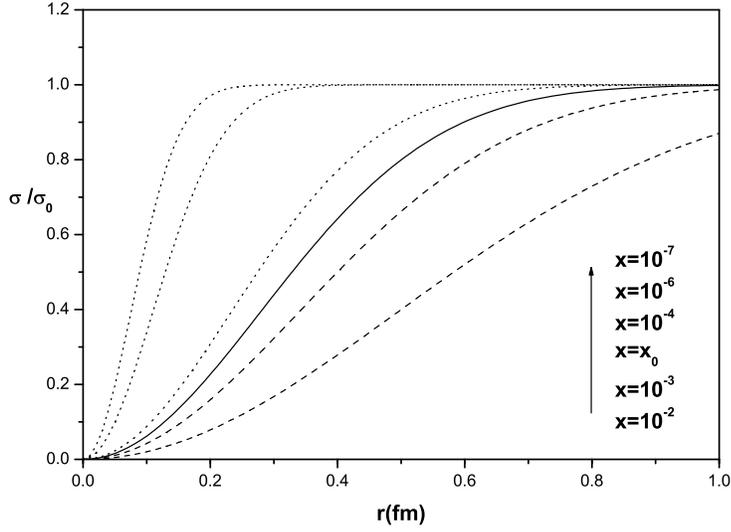}
\caption{The ratio of the dipole cross section with respect to the
effective exponent $\lambda_{s}^{eff}=0.327$ for different values
of $x$.} \label{Fig10}
\end{figure}
\begin{figure}
\centering
\includegraphics[width=1\textwidth]{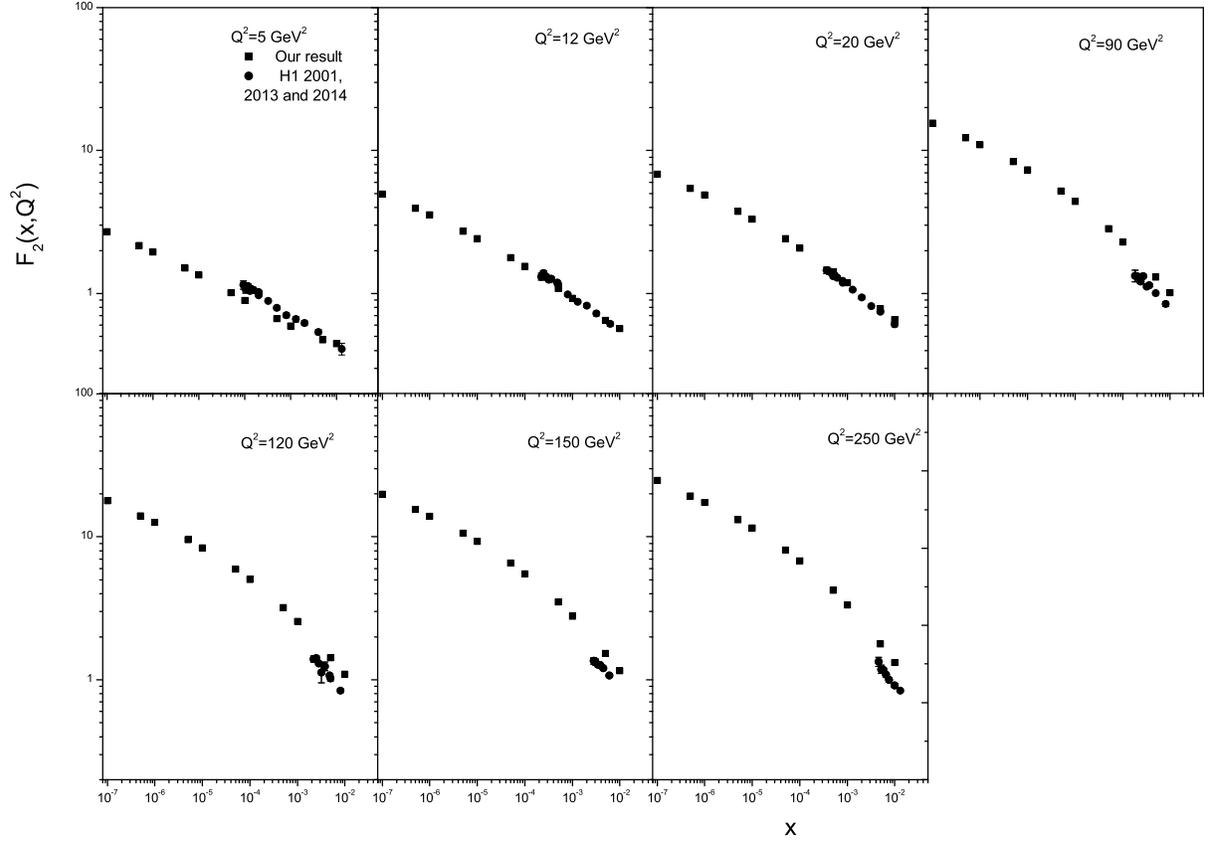}
\caption{Comparison of the proton structure function obtained from
the decoupled evolution equation with the effective exponent with
the H1 data collected since 2001 until 2014 [17-19,22].}
\label{Fig11}
\end{figure}
\begin{figure}
\centering
\includegraphics[width=0.6\textwidth]{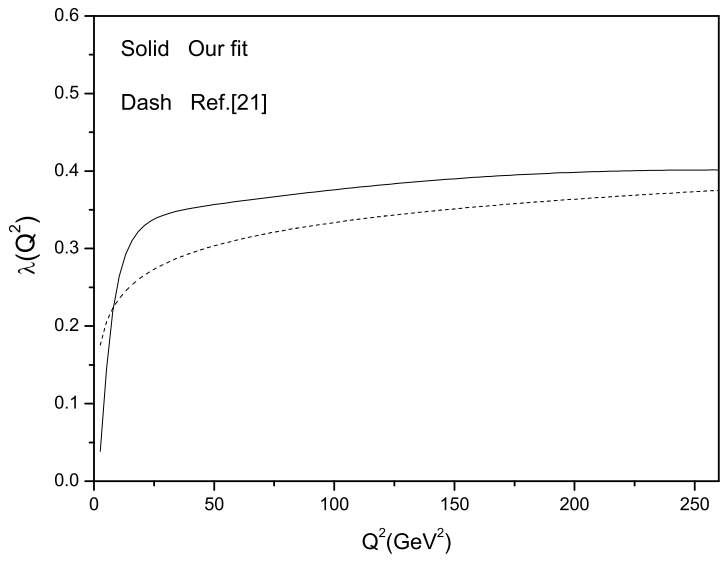}
\caption{Effective exponent $\lambda(Q^{2})$ from averaged
exponents compared with Ref.[21].} \label{Fig11}
\end{figure}
\begin{figure}
\centering
\includegraphics[width=1\textwidth]{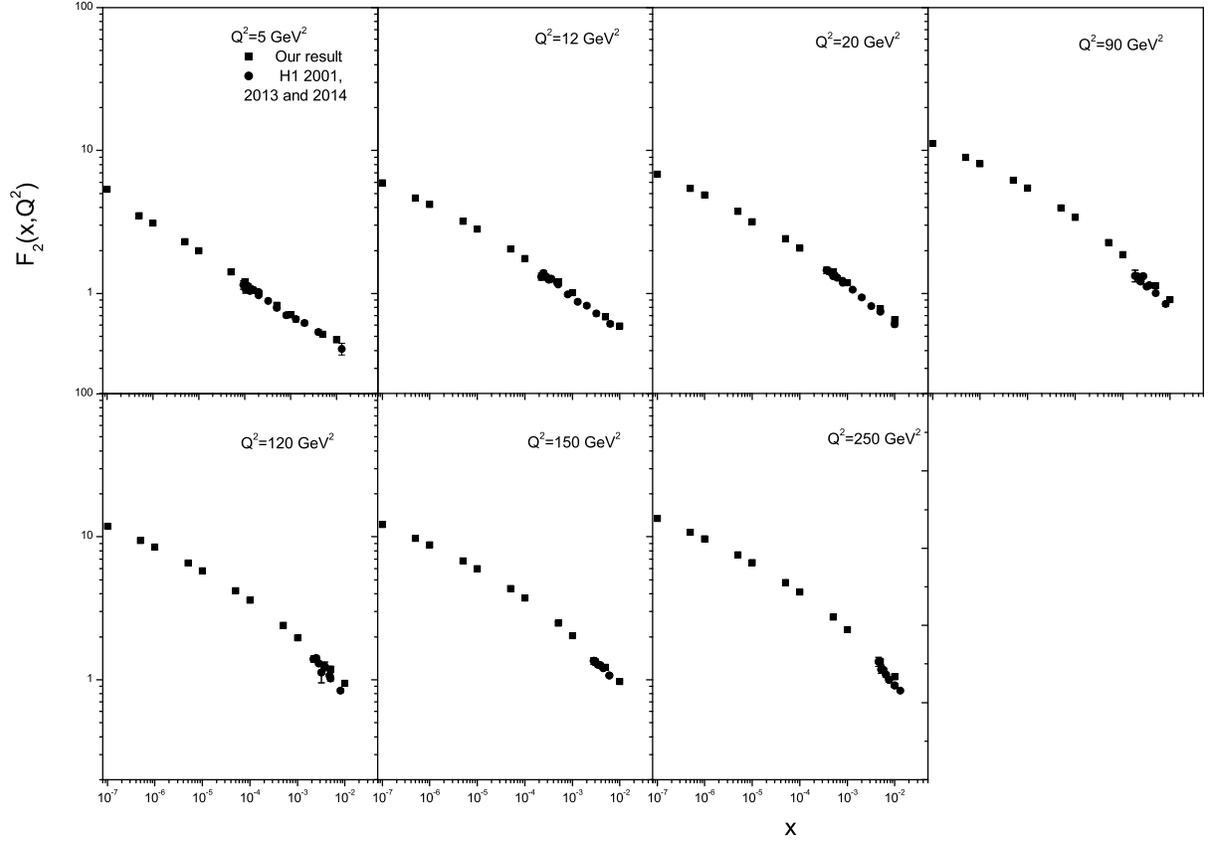}
\caption{The proton structure function obtained by effective
exponent $\lambda(Q^{2})$ compared with H1 data [17-19,22].}
\label{Fig11}
\end{figure}

\newpage
 \begin{table}
\centering \caption{Parameters   determined at $Q^{2}=5~ GeV^{2}$
for regions $x<x_{0}$ and $x>x_{0}$. }\label{table:table1}
\begin{minipage}{\linewidth}
\renewcommand{\thefootnote}{\thempfootnote}
\centering
\begin{tabular}{|l|c|c|c|c|} \hline\noalign{\smallskip} $x$ & $Q_{s}^{2}(GeV^{2})|_{x<x_{0}}$  &
$Q_{s}^{2}(GeV^{2})|_{x>x_{0}}$ &  $R_{0}^{2}(GeV^{-2})|_{x<x_{0}}$ & $R_{0}^{2}(GeV^{-2})|_{x>x_{0}}$  \\
\hline\noalign{\smallskip}
1E-6& 2.160 & -----&  0.463 &-----  \\
5E-6& 1.738 & -----&  0.575 &-----  \\
1E-5& 1.583 & -----&  0.632 &-----  \\
5E-5& 1.274 & -----&  0.785 &-----  \\
1E-4& 1.160 & -----&  0.862 &-----  \\
5E-4& ----- & 0.934& ----- &1.071  \\
1E-3& -----  & 0.850&  ----- &1.177  \\
5E-3& -----  & 0.684&  ----- &1.462  \\
1E-2& -----  & 0.623&  ----- &1.605  \\
\hline\noalign{\smallskip}
\end{tabular}
\end{minipage}
\end{table}
 \begin{table}
\centering \caption{The same as Table III but for $Q^{2}=250~
GeV^{2}$. }\label{table:table1}
\begin{minipage}{\linewidth}
\renewcommand{\thefootnote}{\thempfootnote}
\centering
\begin{tabular}{|l|c|c|c|c|} \hline\noalign{\smallskip} $x$ & $Q_{s}^{2}(GeV^{2})|_{x<x_{0}}$  &
$Q_{s}^{2}(GeV^{2})|_{x>x_{0}}$ & $R_{0}^{2}(GeV^{-2})|_{x<x_{0}}$ & $R_{0}^{2}(GeV^{-2})|_{x>x_{0}}$  \\
\hline\noalign{\smallskip}
1E-6& 7.404 & -----& 0.135 &-----  \\
5E-6& 4.208 & -----&  0.238 &-----  \\
1E-5& 3.300 & -----&  0.303 &-----  \\
5E-5& 1.876 & -----&  0.533 &-----  \\
1E-4& 1.470 & -----&  0.680 &-----  \\
5E-4& -----  & 0.800& ----- &1.255  \\
1E-3& -----  & 0.585&  ----- &1.709  \\
5E-3& -----  & 0.286&  ----- &3.497  \\
1E-2& -----  & 0.210&  ----- &4.761  \\
\hline\noalign{\smallskip}
\end{tabular}
\end{minipage}
\end{table}

\end{document}